# Thermal Brillouin noise observed in silicon optomechanical waveguide

Raphaël Van Laer[1,2], Christopher J. Sarabalis[1], Roel Baets[2], Dries Van Thourhout[2] and Amir H. Safavi-Naeini[1]

E-mail: rvanlaer@stanford.edu

[1] Ginzton Laboratory, Department of Applied Physics, Stanford University, USA
[2] Photonics Research Group, INTEC, Ghent University–imec, Belgium

**Abstract.** Stimulated Brillouin scattering was recently observed in nanoscale silicon waveguides. Surprisingly, thermally-driven photon-phonon conversion in these structures had not yet been reported. Here, we inject an optical probe in a suspended silicon waveguide and measure its phase fluctuatns at the output. We observe mechanical resonances around 8 GHz with a scattering efficiency of $10^{-5}\,\text{m}^{-1}$ and a signal-to-noise ratio of 2. The observations are in agreement with a theory of noise in these waveguides as well as with stimulated measurements. Our scheme may simplify measurements of mechanical signatures in nanoscale waveguides and is a step towards a better grasp of thermal noise in these new continuum optomechanical systems.

PACS numbers: 42.65.Wi, 07.10.Cm, 37.10.Vz, 78.35.+c

## 1. Introduction

The concept of photon-phonon interaction was conceived in the 1920s by Raman, Brillouin and Mandelstam [1–3] in the context of thermal scattering of photons by optical and acoustic phonons. Many previously predicted phenomena became experimentally accessible with the invention of low-loss fibers. In fact, backward Brillouin scattering – a stimulated process initiated by thermal phonons – is a main power limitation in fiber optical networks [4, 5] but has also been used to delay and store light at room temperature [6, 7]. Far afield, Braginsky studied analogous physical processes in gravitational wave detectors – realizing that photon-phonon coupling creates a mechanical instability similar to that in fibers at high optical powers [8]. The field has progressed to observations of photon-phonon conversion in fiber loops [9], photonic crystal fibers [10, 11] and whispering gallery resonators [12–14]. In the last decade, micro- and nano-scale confinement allowed for ever-increasing photon-phonon coupling rates in sub-wavelength fibers [15], chalcogenide rib waveguides [16, 17] and silicon optomechanical crystals [18, 19].

A recent branch of optomechanics seeks to demonstrate efficient coupling between photons and acoustic phonons in nanoscale silicon *waveguides* [20, 21]. These devices require milli- to centimeter interaction lengths and milliwatts of optical input power to observe mechanical action. They possess a continuum of optical and mechanical modes, enabling broadband functionality [22–24] that is otherwise not easily accessible in cavity optomechanics. As the mechanical propagation loss of these waveguides far exceeds their optical propagation loss [25], these waveguides typically provide *optical* gain – often called Brillouin amplification. Within the field of silicon photonics this effect was first seen in silicon/silicon-nitride hybrid suspended waveguides [26] and in silicon pedestal wires [27]. These observations soon led to reconfigurable microwave filters [22, 23]. Cascades of suspended wires enabled gain exceeding the propagation losses last year [28], which was improved recently in suspended silicon rib waveguides [29].

The goal of using these types of coupling to realize quantum and classical information processing, communications and sensing applications has motivated theoretical studies of classical and quantum noise in these structures [24, 30, 31]. Despite the exciting progress, a hallmark of photon-phonon coupling – the thermal Brownian motion – has not yet been observed in these waveguides. Previous work focused either on optical instead of acoustic phonons [32–34], on megahertz vibrations in silicon slots [35, 36] and silica fibers [37, 38] or on gigahertz thermal-initiation of Brillouin amplification in chalcogenide waveguides [39]. In stark contrast, measurements of thermal occupation are standard in cavity optomechanics [40] where the goal is to read out and control the state of a mechanical oscillator. As a step toward understanding noise in these new continuum waveguide systems [24, 25, 30, 31, 41–43], we present the first observation of gigahertz Brownian motion in a silicon waveguide.

## 2. Results

We fabricated a series of silicon wire waveguides through a europractice-ePIXfab multi-project wafer run at imec (fig.1a). The waveguides' thickness was 220 nm and their widths were swept from 450 to 750 nm. The silicon was fully etched in trenches next to most of the waveguide, leaving a silicon core surrounded only by air and silicon dioxide. However, we periodically tapered a thin silicon "socket" layer of thickness 70 nm (fig.1a) next to the silicon core. Finally, we removed the oxide substrate with a buffered hydrofluoric etch (etching for 11 min at 70 nm/min) – letting the socket layer serve as a mask. This produced a cascade of 25 $\mu$m-long suspended silicon beams. The resulting devices are identical to those of [28] except for the anchors, which now consist of the thin silicon socket layer (fig.1a). In our collection of waveguides, we swept the number of suspensions from 30 to 120 at each width. We accessed the wires optically via standard grating couplers [44] for the quasi-TE polarized optical mode (fig.1b).

### *2.1. Experiment*

We injected an optical probe at 1550 nm into the waveguide, aiming to observe Brownian mechanical motion of a Fabry-Pérot-like acoustic mode (fig.1b) that was previously observed by us in stimulated Brillouin measurements [28, 45]. The Brownian motion phase-modulates the optical probe, generating weak red- and blue-shifted sidebands (fig.1c) called the Stokes and anti-Stokes signals. We derived a new theoretical model for the Stokes- and anti-Stokes intensities and spectra in the appendix. A fiber Bragg filter rejected the anti-Stokes signal, leading only probe and Stokes signals into a high-speed photodetector. Thus the optical phase fluctuations were transformed into intensity and photocurrent fluctuations. The probe was pre- and post-amplified by erbium-doped fiber amplifiers to have sufficient power at the photodetector (fig.1c).

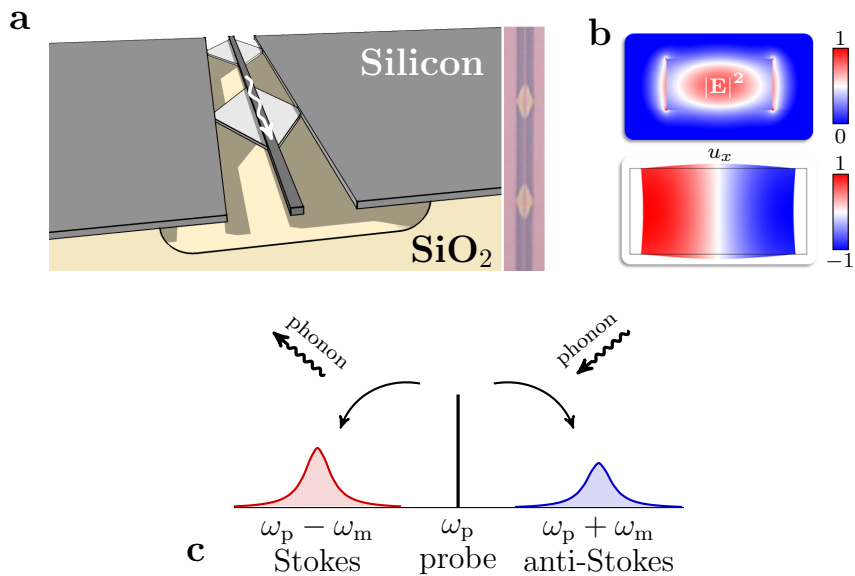

**Figure 1. Thermal Brillouin scattering in suspended silicon waveguide**. **a**, Impression of a silicon-on-insulator waveguide that consists of a series of suspensions and socket anchors (left) and microscope image of a fabricated waveguide (right). The silicon socket layer masks the hydrofluoric etch of the silicon dioxide substrate. The silicon film is 220 nm thick and the wire width varies between 450 and 750 nm. Each suspension is 25 $\mu$m long and the number of suspensions ranges from 30 to 120. Standard grating couplers provide waveguide access at in- and ouput. **b**, Electric field norm of the quasi-TE polarized optical mode (top) and horizontal displacement of the acoustic mode (bottom). **c**, Thermal motion of the waveguide down- and up-converts probe photons of frequency $\omega_\text{p}$ into sidebands of frequency $\omega_\text{p} \pm \omega_\text{m}$ with $\frac{\omega_\text{m}}{2\pi} \approx 8$ GHz.

We observed thermal Brillouin resonances between 5 and 10 GHz with a signal-to-noise ratio of around 2 (fig.3a). Our model (see appendix) predicts the scattering efficiency $\eta$ [1/m] to be

$$\eta = \frac{\omega_\text{p}}{4} \frac{\tilde{\mathcal{G}}}{Q_\text{m}} k_\text{B} T \approx 10^{-5}\,\text{m}^{-1}$$

with $\frac{\omega_\text{p}}{2\pi} = 193$ THz the probe frequency, $\frac{\tilde{\mathcal{G}}}{Q_\text{m}} \approx 12\,\text{W}^{-1}\text{m}^{-1}$ the non-resonant Brillouin gain coefficient measured in stimulated measurements [28, 45] and $k_B T = 25\,\text{meV}$. In other words, in a millimeter of propagation length about 10 probe photons per billion are scattered by thermal phonons. Our model agrees with another recently developed treatment of noise [46] in forward Brillouin interactions.

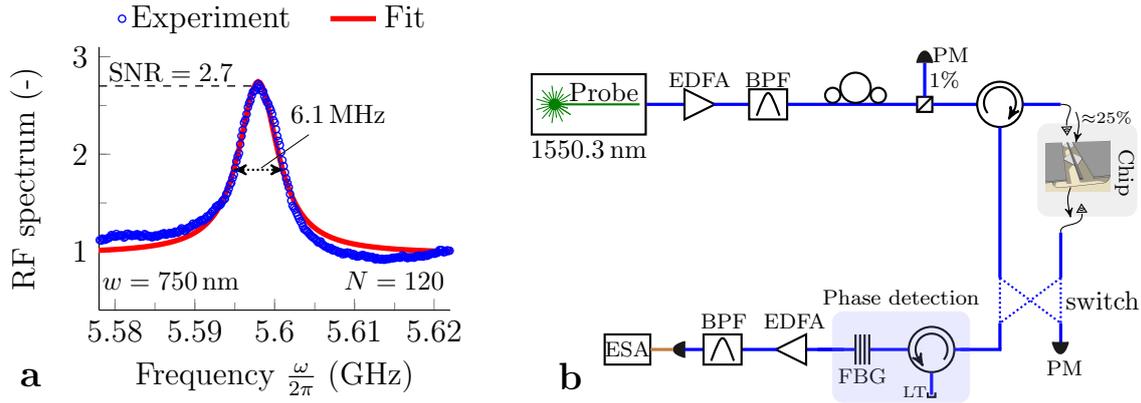

**Figure 2. Thermal Brillouin spectrum in suspended silicon waveguide. a**, An example of a thermal Brillouin resonance in a suspended silicon waveguide of width $w = 750$ nm consisting of $N = 120$ suspensions each $25\,\mu$m long. The resonance has a signal-to-noise ratio of SNR = 2.7 and a quality factor of $Q_\mathrm{m} = 918$. The RF spectrum was normalized with respect to both the shot noise background and the response of the electrical spectrum analyzer (ESA). **b**, We inject an optical probe into the waveguide. It is preamplified by an erbium-doped fiber amplifier (EDFA). The mechanical noise phase-modulates the optical probe as it travels along the waveguide. This optical phase modulation is transducted to intensity modulation by a fiber Bragg grating (FBG) that usually rejects the anti-Stokes signal. We observe identical spectra when the FBG rejects the Stokes instead of the anti-Stokes signal. Thus the Brownian mechanical motion of the waveguide is read out optically on an ESA. A switch determines whether we send the waveguide transmission to the ESA or a power meter (PM). To measure the thermal spectra, the switch is in the "cross" state – leading waveguide transmission to the ESA. The set-up is simpler than gain and cross-phase modulation experiments [26, 28, 29, 45] previously used on such devices.

We ensured that the noise background of the electrical spectrum analyzer scaled with optical power. Assuming this background was set by the erbium-doped fiber amplifer, we obtain (see appendix) an estimated signal-to-noise ratio (SNR) of

$$\mathrm{SNR} = \frac{4\Phi_\mathrm{s}}{F_\mathrm{n}\kappa_\mathrm{m}} \approx 2$$

with $\Phi_\mathrm{s} = \Phi_\mathrm{as} = \eta L \Phi_\mathrm{p}$ the (anti-)Stokes photon flux [1/s], $L$ the suspended waveguide length, $\Phi_\mathrm{p}$ the incoming probe photon flux, $\kappa_\mathrm{m}$ the acoustic decay rate and $F_\mathrm{n}$ the noise figure of the fiber amplifier. Above we inserted typical parameters for our set-up:

$$\hbar\omega_\mathrm{p}\Phi_\mathrm{p} \approx 1\,\mathrm{mW}$$
$$\Phi_\mathrm{p} \approx 10^{16}\,\mathrm{s}^{-1}$$
$$\eta \approx 10^{-5}\,\mathrm{m}^{-1}$$
$$L \approx 1\,\mathrm{mm}$$
$$\frac{\kappa_\mathrm{m}}{2\pi} \approx 10\,\mathrm{MHz}$$
$$F_\mathrm{n} \approx 3$$

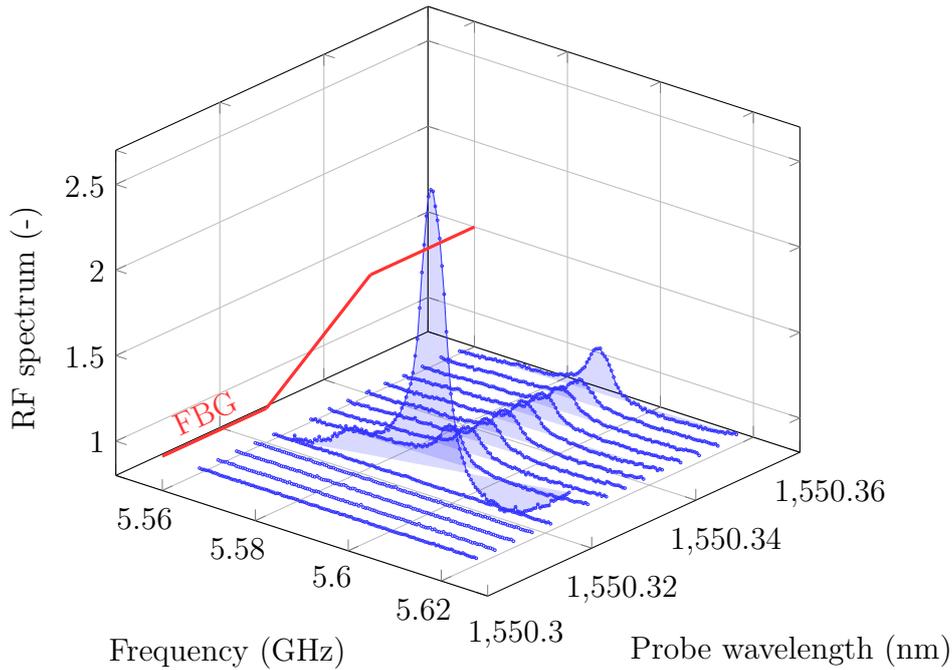

**Figure 3. Thermal Brillouin spectrum as a function of probe wavelength.** The RF spectra depend strongly on the probe wavelength relative to the FBG rejection band. The red curve shows the estimated FBG position as a sketch of transmission (unitless) versus probe wavelength. Within the FBG rejection band the acoustic resonance is not visible at all. As the probe wavelength increases, the probe and the red-shifted Stokes sideband make it through the FBG and next the post-EDFA. Then the interplay between EDFA gain and post-FBG probe power determine the SNR, as long as the blue-shifted anti-Stokes sideband remains rejected by the FBG. We find RF spectra with maximum SNR close to the FBG flank. Figures 2 and 4 are based on the maximum SNR spectra. The FBG's flank is 2.75 GHz (0.22 nm) wide and it has an extinction ratio of 30 dB. This figure concerns a waveguide of width $w = 750$ nm with $N = 120$ suspension, the same one as in fig.2a.

The estimated SNR agrees roughly with the measured signal-to-noise ratios (fig.2a).

Further, the RF spectra depend heavily on the probe wavelength with respect to the fiber Bragg grating's (FBG's) rejection band (fig.3). Within the FBG rejection band, the thermal Brillouin resonances are not visible at all, as too little optical power reaches the photodetector. As the probe wavelength reaches the FBG flank, the probe and its red-shifted Stokes sideband make it through to the photodetector (fig.2b). Then the interplay between the post-EDFA gain – which depends on the probe power – and the post-FBG probe power determine the SNR, as long as the blue-shifted anti-Stokes sideband remains rejected in order to convert phase- to intensity-fluctuations. We find a maximimum SNR close to the FBG flank, and use these spectra to extract the acoustic frequencies $\frac{\omega_\mathrm{m}}{2\pi}$, linewidths $\frac{\kappa_\mathrm{m}}{2\pi}$ and quality factors $Q_\mathrm{m}$ with a Lorentzian fit (fig.3a).

The acoustic resonance frequencies scale inversely with the wires' width (fig.4a), in agreement with finite-element simulations and previous stimulated Brillouin measurements [28,45]. The scaling can be understood if we consider the acoustic mode to be the fundamental excitation of a Fabry-Pérot resonator – with a transverse wavelength equal to twice the waveguide width $w$. This directly implies $\frac{\omega_m}{2\pi} \propto \frac{1}{w}$ [28, 45]. Further, we extract the acoustic quality factors $Q_m$ for a range of wire widths $w$ and number of suspensions $N$. The quality factors are highly dependent on these two parameters, but the pattern is not monotonic. Overall we see higher quality factors in wider waveguides with less suspensions (fig.4b). This trend is likely caused by geometric disorder: the wire width fluctuates along the waveguide, inhomogeneously broadening the acoustic resonance. We saw such effects also in stimulated measurements [28]. A theoretical treatment of inhomogeneous broadening is given in [47]. With a frequency sensitivity of 20 MHz/nm (fig.4a and [28, 45]), the acoustic resonance is vulnerable to even atomic-scale disorder. This is the cost of miniaturization and is a major challenge that must be resolved. Similar effects have been seen in other nanoscale devices, such as silicon rib waveguides [29] and snowflake optomechanical crystals [18].

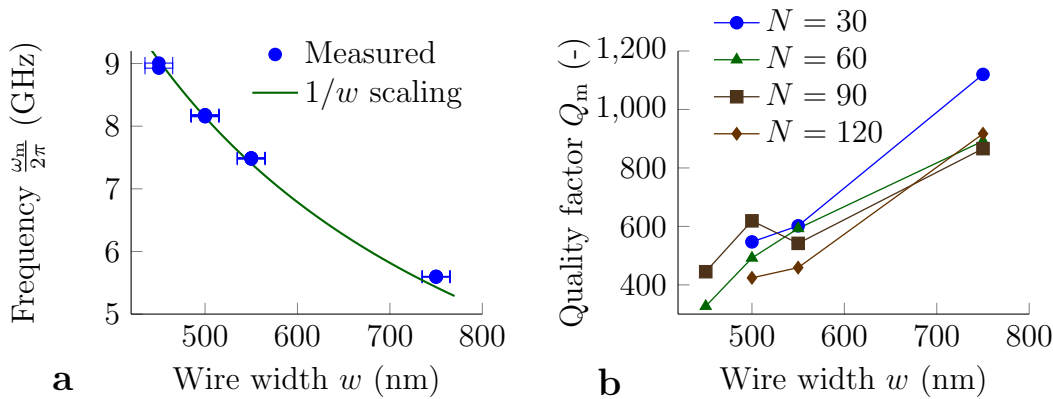

**Figure 4. Mechanical frequencies and quality factors deduced from thermal spectra. a**, The resonance frequencies are in agreement with the $1/w$ Fabry-Pérot model of the acoustic resonance [28, 45]. **b**, The acoustic quality factors are highly dependent on the waveguide width and number of suspensions. The pattern is not monotonic, but generally we find higher quality factors in wider and shorter waveguides. This behavior was also seen in stimulated measurements and is likely caused by geometric disorder along the waveguide [28].

## 3. Conclusion

We observed thermal Brillouin scattering in a silicon photonic waveguide and presented a new model for the scattering efficiencies and noise spectra. Our theory is in exact agreement with another model [46] and in rough agreement with our observations. We analyzed the acoustic frequencies and quality factors for a range of waveguide widths and suspension lengths, finding close correspondence to earlier stimulated Brillouin measurements [28, 45]. The acoustic quality factors are generally higher in shorter and wider waveguides, likely because of atomic-scale geometric disorder. The measurement scheme is simpler than previous stimulated Brillouin set-ups. It is a step towards a better understanding of noise in continuum waveguide systems and may generate alternative approaches to on-chip thermometry [48, 49].

**Appendices**

*Phase-shift model for the thermal spectrum*

We consider a translationally invariant waveguide of length $L$. We inject a weak optical probe at frequency $\frac{\omega_p}{2\pi}$. The probe is weakly phase-modulated by the thermal motion of the waveguide. Assuming the waveguide has a mechanical mode at frequency $\frac{\omega_m}{2\pi}$, this generates sidebands at angular frequency $\omega_p \pm \omega_m$. Here we seek the spectrum and intensity of the scattered light – neglecting optical driving of the mechanical mode. We denote the optical field at the output of the waveguide by $a_{\text{out}}(\tau)$. The mechanical motion phase-shifts the input field $a_p(\tau)$:

$$a_{\text{out}}(\tau) = a_p(\tau)e^{i\phi(\tau)} + \xi(\tau) \approx a_p(\tau)(1 + i\phi(\tau)) + \xi(\tau)$$

with $\phi(\tau)$ a random variable with vanishing ensemble average ($\langle\phi(\tau)\rangle = 0$) and $\xi(\tau)$ the vacuum noise. Here we treat the phase shift $\phi$ as induced by a lumped element. This is a good approximation since in forward intra-modal Brillouin scattering momentum conservation dictates that the acoustic phase velocity equals the optical group velocity [26, 45]. Thus $\Lambda = \frac{2\pi v_g}{\omega_m} \approx 1\,\text{cm} \gg L \approx 1\,\text{mm}$ with $\Lambda$ the acoustic wavelength and $v_g = 7 \cdot 10^7\,\text{m/s}$ the optical group velocity. Therefore the finite acoustic wavevector can be considered zero.

*Field spectral density.* First we look at the spectrum of the output *field*, which would result from a spectrally selective photon-counting experiment. The autocorrelation of output field is

$$\langle a_{\text{out}}^\dagger(\tau)a_{\text{out}}(0)\rangle = \langle a_p^\dagger(\tau)a_p(0)\rangle + \sqrt{\Phi_p}\langle\phi(\tau)\phi(0)\rangle + \langle\xi^\dagger(\tau)\xi(0)\rangle$$

with $\Phi_p$ the probe flux [1/s] and where we treated the phase shift $\phi$ and the probe field $a_p$ as uncorrelated and $\phi$ as a real observable. We took $a_p(\tau) = \sqrt{\Phi_p}$ to be a constant in the second term, neglecting noise in the probe. The output field spectral density $S_a(\omega)$ equals

$$S_a(\omega) = \int_{-\infty}^{\infty} d\tau\, e^{i\omega\tau}\langle a_{\text{out}}^\dagger(\tau)a_{\text{out}}(0)\rangle$$
$$= S_p(\omega) + \Phi_p S_\phi(\omega)$$

with $S_p(\omega)$ the probe spectrum, $S_\phi(\omega)$ the spectral density of the phase and we used $\langle\xi^\dagger(\tau)\xi(0)\rangle = 0$ for optical frequencies with negligible thermal occupation. We also have

$$\phi(\tau) = k_p\, \partial_q n_{\text{eff}}\, q(\tau) L$$

with $k_\mathrm{p} = \frac{\omega_\mathrm{p}}{c}$ the vacuum probe wavevector, $c$ the speed of light, $\partial_q n_\mathrm{eff}$ the sensitivity of the optical effective index $n_\mathrm{eff}$ to motion $q$ and $q(\tau)$ a coordinate describing the average motion along the entire waveguide of length $L$. Its ensemble average vanishes ($\langle q(\tau)\rangle = 0$) and it can be written as $q(\tau) = \frac{1}{L}\int_0^L q(z,\tau)dz$ with $q(z,t)$ the actual motion at point $z$ along the waveguide. Note that $q(\tau)$ represents a single degree of freedom if the curvature of the phonon band is neglected – an excellent approximation here.‡ It is therefore an effective harmonic oscillator that corresponds to the average motion along the entire waveguide. Thus the spectral density of the phase is

$$S_\phi(\omega) = \int_{-\infty}^\infty d\tau\, e^{i\omega\tau} \langle \phi(\tau)\phi(0)\rangle$$
$$= (k_\mathrm{p}\partial_q n_\mathrm{eff} L)^2 S_q(\omega)$$

Further, the spectral density of the motion $S_q(\omega)$ is [50]

$$S_q(\omega) = q_\mathrm{zpf}^2 \left( \frac{\kappa_\mathrm{m}\overline{n}}{(\omega_\mathrm{m}+\omega)^2 + \kappa_\mathrm{m}^2/4} + \frac{\kappa_\mathrm{m}(\overline{n}+1)}{(\omega_\mathrm{m}-\omega)^2 + \kappa_\mathrm{m}^2/4} \right)$$

in thermal equilibrium for a weakly damped oscillator and with $q_\mathrm{zpf} = \sqrt{\frac{\hbar}{2m_\mathrm{eff}L\omega_\mathrm{m}}}$ the zero-point fluctuation, $m_\mathrm{eff}$ the effective modal mass per unit length and $\overline{n}$ the thermal phonon occupation in the degree of freedom described by $q(\tau)$. At 300 K we have $\overline{n} \approx \frac{k_\mathrm{B}T}{\hbar\omega_\mathrm{m}} = 624$ for a 10 GHz acoustic mode. This implies the two sidebands are of nearly equal intensity. Besides, $\overline{n}$ is the number of phonons in the degree of freedom captured by the coordinate $q(\tau)$ as defined above. This phonon occupation $\overline{n}$ does not scale with waveguide length $L$, which means that $S_q(\omega) \propto \frac{1}{L}$ § and therefore $S_\phi(\omega) \propto L$ as expected from a random walk: the root-mean-square of the phase $\sqrt{\langle\phi^2\rangle}$ goes as $\sqrt{L}$. Thus the blue- and red-shifted sideband intensities increase linearly with length.

Previously we showed that [45]

$$\tilde{\mathcal{G}} = 2\omega_\mathrm{p}\frac{Q_\mathrm{m}}{k_\mathrm{eff}}\left(\frac{1}{c}\partial_q n_\mathrm{eff}\right)^2$$

with $k_\mathrm{eff} = m_\mathrm{eff}\omega_\mathrm{m}^2$ the effective mechanical stiffness per unit length, $\tilde{\mathcal{G}}$ the Brillouin gain coefficient [1/(Wm)] and $Q_\mathrm{m} = \frac{\omega_\mathrm{m}}{\kappa_\mathrm{m}}$. So the total output spectral density of the field is

$$S_a(\omega) = S_\mathrm{p}(\omega) + \Phi_\mathrm{p}L\eta\left(\frac{\kappa_\mathrm{m}}{(\omega_\mathrm{m}+\omega)^2 + \kappa_\mathrm{m}^2/4} + \frac{\kappa_\mathrm{m}}{(\omega_\mathrm{m}-\omega)^2 + \kappa_\mathrm{m}^2/4}\right)$$

‡ Consider the waveguide of length $L$ as a series of independent oscillators. The phase shift then becomes $\phi \propto \sum_z q_z$ with $q_z$ the motion of the oscillator at position $z$. Applying periodic boundary conditions and with Fourier decomposition $q_k = \frac{1}{\sqrt{L}}\sum_z e^{-ikz}q_z$, we see that $\phi \propto q_{k=0}$, representing a single degree of freedom that appears quadratically in the energy. If the phonon band is flat, the precise boundary conditions do not impact this conclusion. The band of the mechanical mode studied here remains flat if the acoustic wavelength is longer than just a couple of microns.
§ The scaling $S_q(\omega) \propto \frac{1}{L}$ can be understood from $\langle q(z',\tau)q(z,\tau)\rangle \propto \delta(z'-z)$: the thermal Brownian motion is delta-correlated along the waveguide, unlike in stimulated Brillouin schemes with quadratic build-up [45, 46].

with $\eta$ the scattering efficiency [1/m] and

$$\eta = \frac{\omega_\text{p}}{4} \frac{\tilde{\mathcal{G}}}{Q_\text{m}} k_\text{B} T$$

This result is in agreement with another recently developed noise model [46] for forward Brillouin interactions. Since

$$\int_{-\infty}^{\infty} \frac{d\omega}{2\pi} \frac{\kappa_\text{m}}{(\omega_\text{m} \pm \omega)^2 + \kappa_\text{m}^2/4} = 1$$

we have

$$\Phi_\text{s} = \Phi_\text{as} = \Phi_\text{p} L \eta$$

with $\Phi_\text{s}$ and $\Phi_\text{as}$ the Stokes and anti-Stokes photon flux.

*Current spectral density.* Instead of doing a photon-counting experiment, we use a high-speed photodetector. Photodetectors cannot directly detect the phase $\phi(\tau)$. Therefore, we first send the output field through a fiber Bragg filter, rejecting either the Stokes or anti-Stokes sideband. Thus the symmetry between the sidebands is broken and the Stokes (or anti-Stokes) signal produces a beat note at $\omega_\text{m}$ in the photocurrent $I(\tau)$. Here we briefly discuss the properties of the current spectral density. The filtered output field is

$$a_\text{out}(\tau) = a_\text{pf}(\tau) + a_\text{s}(\tau) + \xi(\tau)$$

with the Stokes (or anti-Stokes) field $a_\text{s}(\tau) = i\sqrt{\Phi_\text{p}}\phi_\text{f}(\tau)$ and where "f" stands for the filtered fields after the fiber Bragg grating. Note that the field spectrum is asymmetric because of the filter. The current $I(\tau)$ is

$$I(\tau) = a_\text{out}^\dagger(\tau) a_\text{out}(\tau) = I_\text{pf}(\tau) + \sqrt{\Phi_\text{p}} \left( a_\text{s}(\tau) + a_\text{s}^\dagger(\tau) + \xi(\tau) + \xi^\dagger(\tau) \right)$$

with $I_\text{pf}(\tau)$ the probe current and where we neglected weak currents associated with the Stokes field and the shot noise. So the current spectral density $S_{II}(\omega)$ is

$$S_{II}(\omega) = \int_{-\infty}^{\infty} d\tau \, e^{i\omega\tau} \langle I(\tau) I(0) \rangle$$
$$= S_{I_\text{pf}}(\omega) + \Phi_\text{p}^2 \left( S_{\phi_\text{f}}(\omega) + S_{\phi_\text{f}}(-\omega) \right) + F_\text{n} \Phi_\text{p}$$

where we used $\langle \xi(\tau) \xi^\dagger(0) \rangle = \delta(\tau)$. We multiplied the shot noise (third term) by the noise figure $F_\text{n} \approx 3$ of the EDFA. During the experiment, we ensured that the background on the electrical spectrum analyzer scaled with the optical power. Further, the spectral density $S_{\phi_\text{f}}(\omega)$ of the filtered phase $\phi_\text{f}$ is

$$S_{\phi_\text{f}}(\omega) = L\eta \frac{\kappa_\text{m}}{(\omega_\text{m} \pm \omega)^2 + \kappa_\text{m}^2/4}$$

where the $\pm$ indicates whether the fiber Bragg filter rejected the Stokes ($+$) or anti-Stokes ($-$) – which is determined by the probe wavelength. Assuming the EDFA sets the background, the signal-to-noise ratio equals

$$\text{SNR} = \frac{\Phi_\text{p} S_{\phi_\text{f}}(\omega_\text{m})}{F_\text{n}} = \frac{4\Phi_\text{s}}{F_\text{n} \kappa_\text{m}}$$

**Acknowledgement**

R.V.L. acknowledges the Agency for Innovation by Science and Technology in Flanders (IWT) for a PhD grant, FWO for a travel grant and VOCATIO for a fellowship. This work was also supported by NSF ECCS-1509107.